\documentclass[preprint2]{aastex}

\usepackage{epsf}

\begin{document}

\title {Hydra A at Low Radio Frequencies}

\author{W.M. Lane}

\affil{Naval Research Lab, Code 7213, 4555 Overlook Ave. SW,
Washington, DC, 20375}
\email{wendy.lane@nrl.navy.mil}

\author{T.E. Clarke} 
\affil{Department of Astronomy, University of Virginia, P.O. Box 3818,
Charlottesville, VA 22903-0818 }
\email{tclarke@virginia.edu}

\author{G.B. Taylor}
\affil{National Radio Astronomy Observatory, P.O. Box O, Socorro, NM
87801}
\email{gtaylor@nrao.edu}

\author{R.A. Perley}
\affil{National Radio Astronomy Observatory, P.O. Box O, Socorro, NM
87801}
\email{rperley@nrao.edu}

\and

\author{N.E. Kassim}
\affil{Naval Research Lab, Code 7213, 4555 Overlook Ave. SW,
Washington, DC, 20375}
\email{namir.kassim@nrl.navy.mil}

\begin{abstract}  

We present new, low-frequency images of the powerful FR I radio galaxy
Hydra A (\objectname[]{3C 218}).  Images were made with the Very Large
Array (VLA)\footnote{operated by The National Radio Astronomy
Observatory, a facility of the National Science Foundation operated
under cooperative agreement by Associated Universities, Inc.} at
frequencies of $1415$, $330$, and $74$ MHz, with resolutions on the
order of $20\arcsec$.  The morphology of the source is seen to be more
complex and even larger than previously known, and extends nearly
$8\arcmin$ (530 kpc) in a North-South direction. The southern lobe is
bent to the east and extends in that direction for nearly $3\arcmin$
(200 kpc).  In addition, we find that the northern lobe has a flatter
spectral slope than the southern lobe, consistent with the appearance
of greater confinement to the south.  We measure overall spectral
indices $\alpha^{330}_{74} = -0.83$ and $\alpha^{1415}_{330} = -0.89$.
\end{abstract}


\section{Introduction}

Hydra A (3C 218) is a high luminosity Fanaroff-Riley type I (FR I)
radio galaxy.  Existing radio maps at frequencies greater than 5 GHz
have mapped only the central, bright portions of the source. The inner
jets extend for roughly $1.5\arcmin$ along an axis at $26\degr$, and
exhibit a high degree of ``S'' symmetry, which Taylor et al. (1990)
suggest may be due to precession of the central engine.  Very Long
Baseline Array (VLBA) maps at 1.3, 5, and 15 GHz also show a highly
symmetric source on parsec scales, while VLBA measurements of HI
absorption against the core suggest an HI disk with a height of
$\sim30 h^{-1}_{50}$ pc and a rotation axis which is aligned with the
jet axis (Taylor 1996).  Measurements with the Very Large Array (VLA)
did not find any HI absorption against the kpc-scale jets
(Dwarakanath, Owen, \& van Gorkom 1995).  The source has been mapped
with low angular resolution ($\sim 45$\arcsec) at 1500 MHz (Taylor et
al. 1990), revealing a far more extended structure which covers nearly
$8\arcmin$ in a north-south direction, but obscuring its details.
Hydra A has not been studied in detail at any lower radio frequencies.

The host system is identified as a cD galaxy (Matthews, Morgan, \&
Schmidt 1964) with a double optical nucleus (Ekers \& Simkin 1983).
It is the dominant member of the poor Abell cluster A780 (Abell 1958),
and lies at a redshift $z = 0.0542$ (Dwarakanath et al. 1995; Owen,
Ledlow \& Keel 1995; Taylor 1996).  A central gas disk, as measured by
optical emission lines, has a rotation axis of $29\degr \pm 9\degr$
(eg. Simkin 1979), in good agreement with the kpc-scale radio jet
axis.

The cluster is a strong X-ray source and has been classified as a
cooling flow cluster with a mass accretion rate of $300$ M$_\odot$
yr$^{-1}$ (David et al. 2001).  Using data from the Chandra telescope,
Sambruna et al. (2000) find a point source which is spatially
coincident with the radio core, embedded in a diffuse X-ray halo.
They interpret this as evidence for a low luminosity active galactic
nucleus (AGN), with most of its optical/UV emission obscured by
intrinsic reddening.

The high symmetry and near-unity jet-counterjet ratios of the core
structures in the VLBA maps suggest that Hydra A is aligned within a
few degrees of the plane of the sky on parsec scales (Taylor 1996).
However a strong Faraday rotation measure (RM) and asymmetry in both
the RM and depolarization between the kpc-scale jets suggest an
inclination angle of $48\degr$ (Taylor \& Perley 1993), with the
northern jet lying closer to us.  The difference in these measurements
may be explained if the jet gas either undergoes a sharp bend or is
moving non-relativistically.  

Here we present moderate resolution maps of Hydra A at three low-radio
frequencies: $1415$ MHz, $330$ MHz, and $74$ MHz.  Throughout this
paper we use H$_o = 65$ km s$^{-1}$ Mpc$^{-1}$, and define spectral
index $\alpha$ such that S$_{\nu} \propto \nu^{\alpha}$.
 
\section{Radio Data}

\begin{figure}
\epsscale{0.80}
\plotone{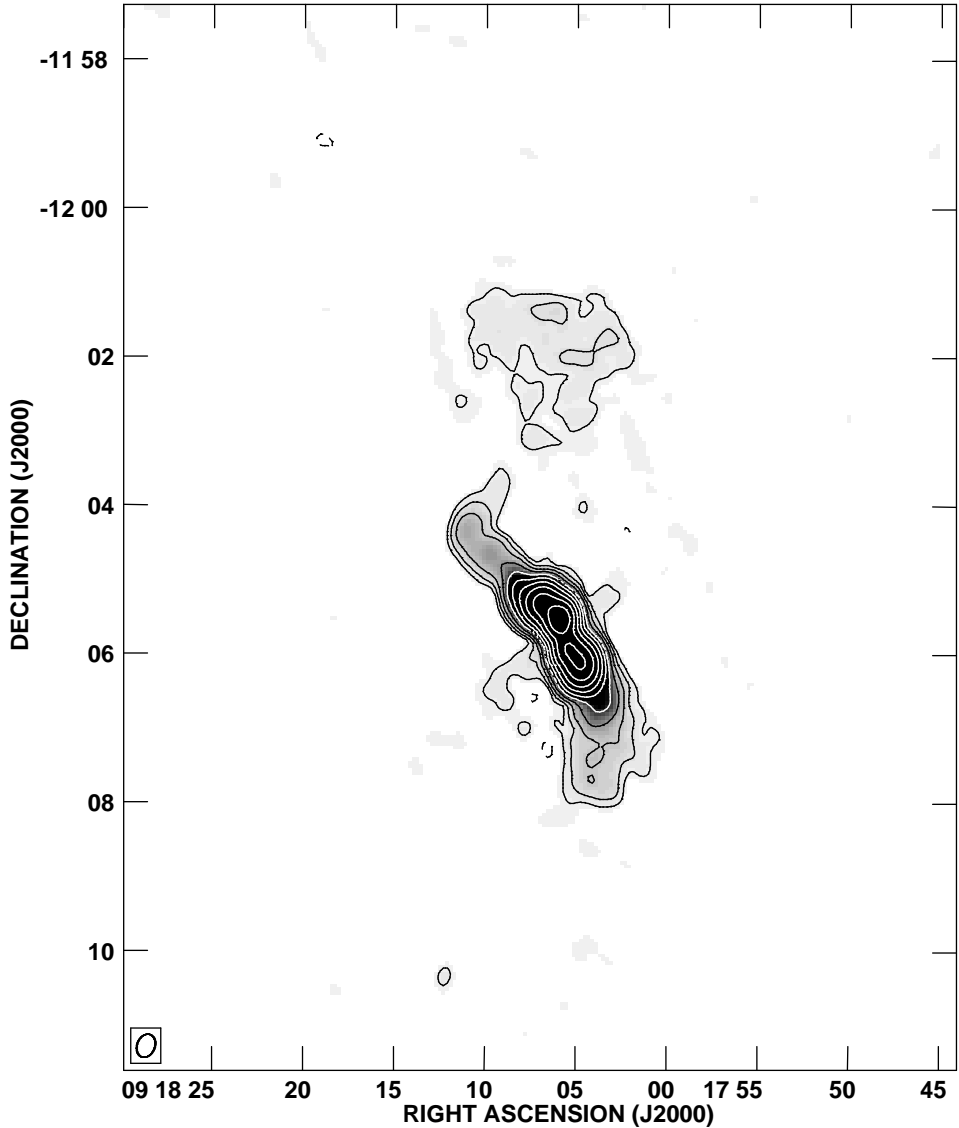}
\caption[lane.fig1.ps]{Contour image of Hydra A at 1415
MHz, using data from the C and D configurations of the VLA.  Note the
central bright core and jets typically seen in higher frequency images
of this source.  In addition we are just able to see the northern jet
bending and then expanding into a diffuse lobe.  The extensions to the
east and west of the core are imaging artifacts, and extend throughout
the image as a low level stripe.  The dynamic range in the image is
5500:1, and the noise level is dynamic range limited.  The peak is
11.8 Jy bm$^{-1}$.  The contours are at multiples of the $5\sigma$
noise level, $0.0126 \times (-1, 1, 2, 4, 8, 16 ..)$ Jy bm$^{-1}$. The
clean beam is $19\farcs3 \times 14\farcs3$ at a position angle of
$-21\fdg4$, and the image has been corrected for primary beam
attenuation. \label{fig:HydL}}
\end{figure}

Hydra A was observed in the C- and D-configurations of the VLA at
frequencies near 1415 MHz in 1988/1989 (see Table ~\ref{tab:Obs}).
The observations were made in a continuum mode, with bandwidths of
12.5 MHz.  The original goal of these observations was to study the
polarization properties of the source.  We have re-reduced and imaged
the data using standard routines in the Astronomical Image Processing
System (AIPS).  The majority of the flux at this frequency lies in the
inner jets.  In order to better map the lower flux density features of
the weak extended lobes, we improved our dynamic range with a baseline
correction, which we calculated using the observed primary flux
calibrators 3C~286 and 3C~147.  The final image (see Figure
~\ref{fig:HydL}) has a resolution of $19\farcs3 \times 14\farcs3$ at a
position angle of $-21\fdg4$.  The noise in the map remains limited by
the dynamic range, which is 5500:1.  In addition, incomplete
uv-coverage has led to noise artifacts near the source.

\begin{figure}
\epsscale{0.80}
\plotone{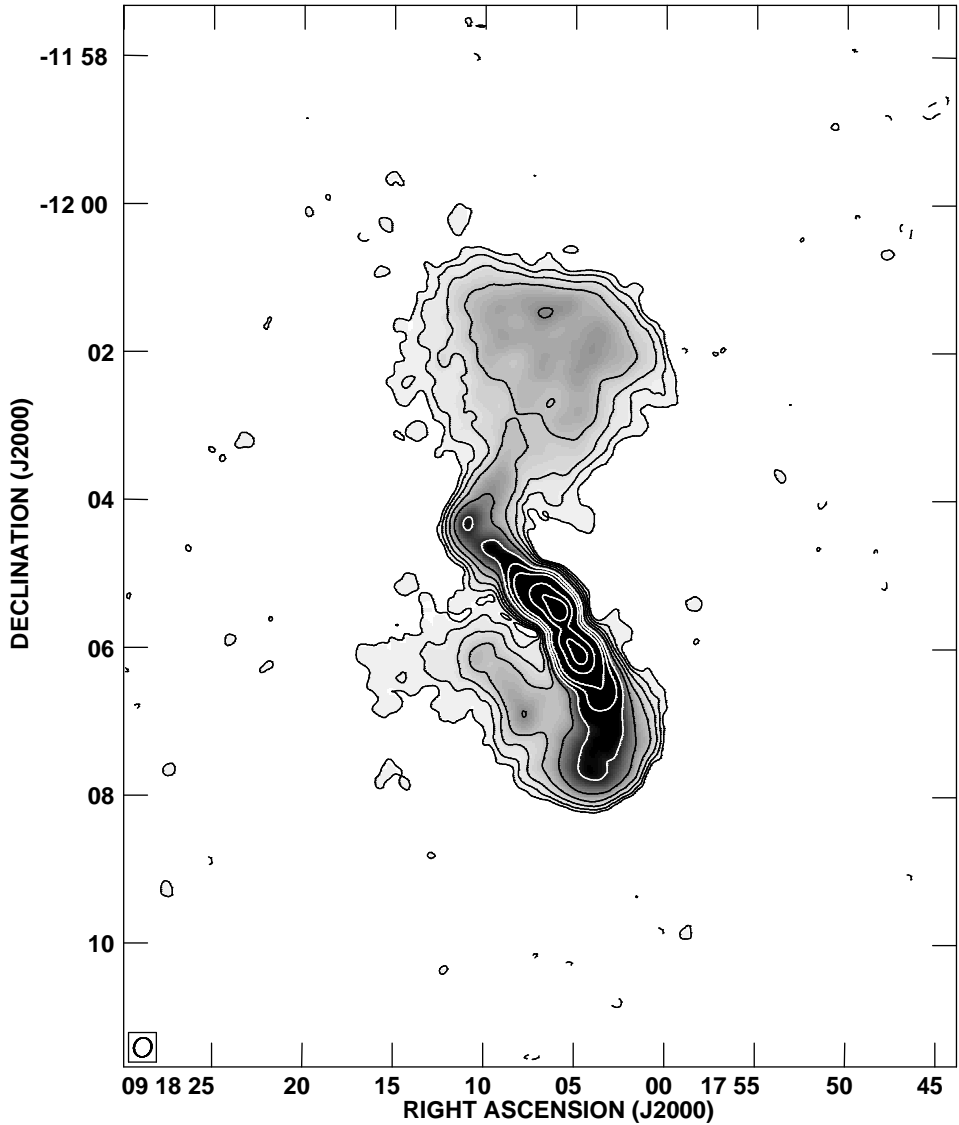}
\caption[lane.fig2.ps]{Contour image of Hydra A at 330 MHz,
using combined data from the VLA in A, B, and C-configurations.  Note
the extended structure here, with the central core and bright jets
extending further and turning sharply before expanding into diffuse
lobes. The dynamic range in the image is 5000:1, and the clean beam is
$15\farcs7 \times 13\farcs9$ at a position angle of $-21\fdg5$.  The
peak flux is 27.76 Jy bm$^{-1}$. The contours are at multiples of the
$3\sigma$ noise level, $0.01605 \times (-1, 1, 2, 4, 8, 16 ...)$ Jy
bm$^{-1}$.  This image has been corrected for the shape of the primary
beam.\label{fig:HydP}}
\end{figure}

Short observations in the A-, B-, and C- configurations of the VLA
were carried out at frequencies near 330 MHz on various dates between
1990 and 1998 (see Table ~\ref{tab:Obs}).  Most of these data were
intended to provide phase-referencing information for simultaneous
observations at 74 MHz.  The observations were set up as multiple
short integrations spread out over a wide hour angle range to provide
good uv-coverage.  Standard calibration observations were used.  The
data were mapped in AIPS using wide-field imaging techniques, which
correct for distortions in the image caused by the non-coplanarity of
the VLA over a wide field of view (the 3-D effect) by using a set of
small overlapping maps, or ``facets'' to cover the desired image area
(Cornwell \& Perley 1992).  The rms noise in the final image is
limited by the dynamic range of 5000:1 (see Figure ~\ref{fig:HydP}).
The dataset has a maximum resolution of $8\arcsec$. In order to more
clearly show the extended lobe emission we have chosen a display
resolution of about $15\arcsec$.

\begin{figure}
\epsscale{0.80}
\plotone{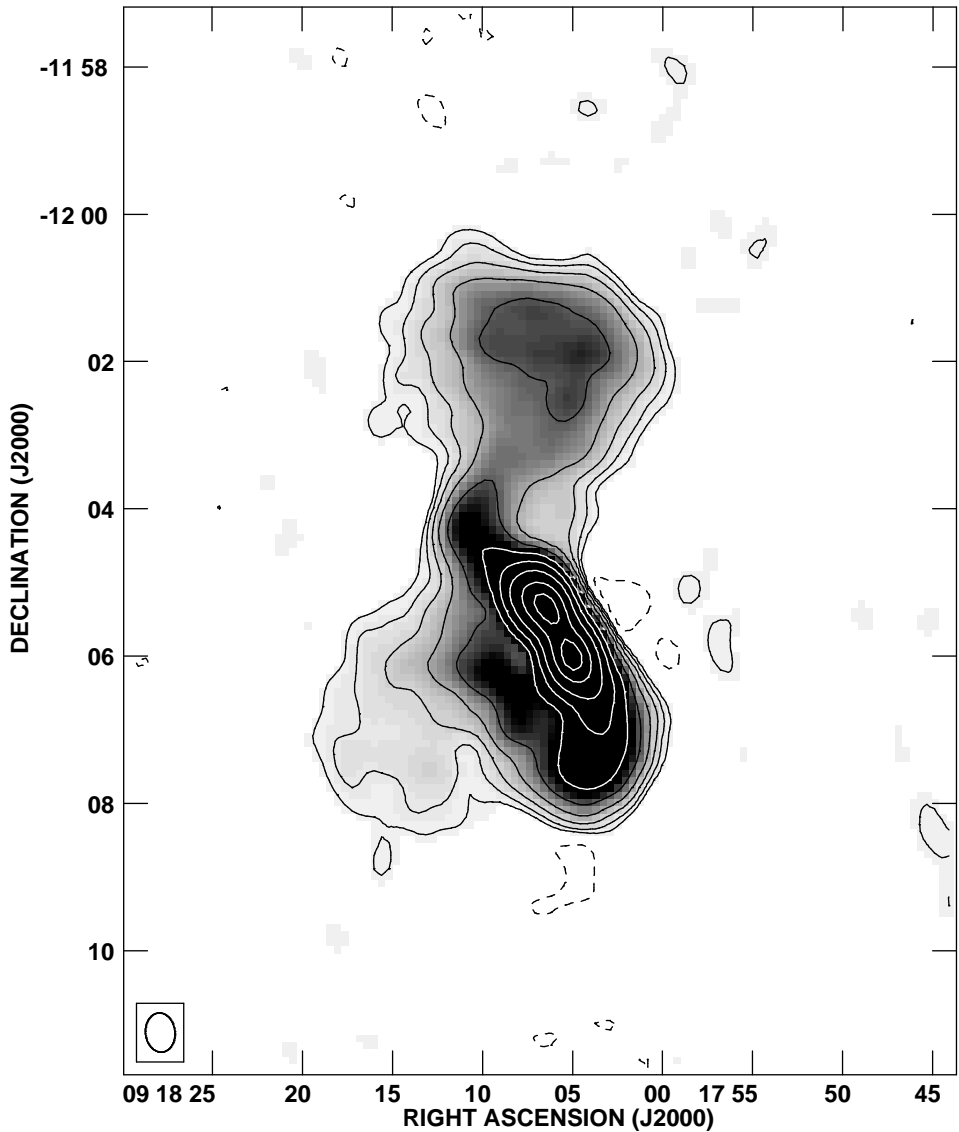}
\caption[lane.fig3.ps]{Contour image of Hydra A at 74 MHz,
using data from the VLA in A configuration.  Note that the southern
lobe appears to be slightly more extended than in the 330 MHz map.
The dynamic range in the image is 1450:1, and the clean beam is
$31\farcs9 \times 23\farcs7$ at a position angle of $6\fdg4$. The peak
flux is 96.95 Jy bm$^{-1}$. The contours are at multiples of the
$3\sigma$ noise level, $0.1993 \times (-1, 1, 2, 4, 8, 16, ...)$ Jy
bm$^{-1}$.  This map has been corrected for the attenuation of the
primary beam.\label{fig:Hyd4}}
\end{figure}

Observations at 74 MHz were made in the A-configuration of the VLA on
2002 April 29 for 4.2 hours. These data were taken as part of a
program to observe the galaxy cluster Abell 754 which is located
$3\fdg3$ from Hydra A, well within the primary beam of the antennae at
74 MHz, which have a full-width at half maximum of $11\fdg7$. The data
were taken in spectral-line mode to help with radio frequency
interference excision, and the total bandwidth is 1.56
MHz. Observations and models of Cygnus A were used to correct the
bandpass and set the antenna gains.  The data were averaged to a
channel resolution of 97.7 kHz and imaged using the wide-field imaging
technique described above. One of the individual facets was centered
on Hydra A: it is this image we present in Figure ~\ref{fig:Hyd4}.
The dynamic range in the image is 1450:1, and the rms noise after
correcting for the attenuation of the primary beam is 67 mJy
bm$^{-1}$. The angular resolution is $31\farcs9 \times 23\farcs7$ at a
position angle of $6\fdg4$.

Maps at all three frequencies do not resolve the core, but show the
inner bright jets extending north and south to cover $\sim 1.5\arcmin$
(100 kpc), and longer, outer jets extending $4\arcmin$ (265 kpc). The
jets are then seen to bend sharply and expand into diffuse, extended
lobes.  The northern lobe is seen at all three radio frequencies,
while the southern lobe is visible only at the two lower frequencies.
The entire structure covers roughly $8\arcmin$ (530 kpc) in a
north-south direction, and displays an ``S''-shaped appearance.

Previous higher frequency observations did not detect the extended
diffuse radio lobes in this source (eg. Taylor et al. 1990). This is
not completely a surprise given that the radiative lifetime of the
particles is longer at lower frequencies.  Although there is currently
insufficient data at appropriate sensitivity and resolution
combinations to determine if such extended structure is typical of
FR1's as a class, it has been observed in a few other objects (eg.,
M87; Owen, Eilek \& Kassim 2000).

\section{Spectral Index}

\begin{figure}
\plottwo{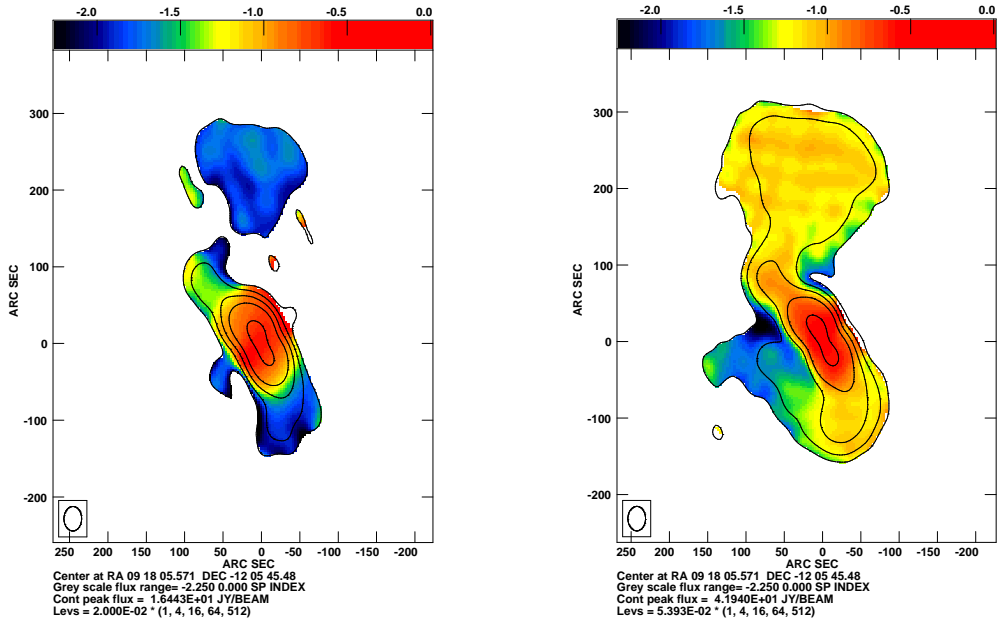}{lane.fig4b.ps}
\caption[lane.fig4a.ps, lane.fig4b.ps]{{\it left}:
Spectral index map of Hydra A between 1415 MHz and 330 MHz is shown in
color, with 1415 MHz contours overlaid. The input maps were clipped at
$5\sigma$ to create this spectral index image, which has a resolution
of $32\arcsec \times 23\arcsec$.  The contours are at multiples of the
$5\times$ the rms noise level in the 1415 MHz image, $0.020 \times
(1, 4, 16, 64, 512)$ Jy bm$^{-1}$.  Errors in the spectral index at a
$1\sigma$ level range from $\pm 0.03$ at the core, to $\pm 0.04$ in
the jets and $\pm 0.08$ in the northern lobe.  {\it right}: Spectral
index map of Hydra A between 330 MHz and 74 MHz is shown in color,
with 330 MHz contours overlaid.  The input maps were clipped at
$5\sigma$ to create this spectral index image, which has a resolution
of $32\arcsec \times 23\arcsec$.  The contours are at multiples of
$5\times$ the rms noise level in the 330 MHz image, $0.05393 \times
(1, 4, 16, 64, 512)$. Errors in the spectral index at a $1\sigma$
level are $\pm 0.03$ near the core and through the jets, $\pm 0.04$ in
the northern lobe and as high as $\pm 0.1$ at the end of the southern
lobe. \label{fig:HydSI}}
\end{figure}

In order to enable a direct comparison between data at different
frequencies, the UV-data were limited in UV-range to $0.13 < UV < 9$,
and tapered to match the dirty beams as closely as possible at each
frequency.  An identical clean beam of $23\arcsec \times 32\arcsec$ at
a position angle of $0\degr$ was used to restore the maps.  The images
were then aligned, clipped at a $5\sigma$ level, and combined to
create spectral index images (see Figure ~\ref{fig:HydSI}).  Errors
were estimated based on a combination of the signal to noise ratio at
a given pixel, and the assumption of a $3\%$ amplitude calibration
accuracy at each frequency.  We assume that at these frequencies there
are no significant flux density variations in the source over the 14
years spanned by the data.

The spectral index between 1415 and 330 MHz is $\alpha\sim-0.60 \pm
0.03$ near the core of the source.  The northern jet spectral index
changes slowly to $\alpha\sim-1.50 \pm 0.04$ at $125\arcsec$ and then
reaches $\alpha \sim -1.70 \pm 0.05$ as it turns and broadens into the
northern lobe.  In contrast, the southern jet slope steepens quickly
from $\alpha \sim -1.10 \pm 0.03$ at $40\arcsec$, to $\alpha \sim -1.90 \pm
0.05$ at $125\arcsec$.  The southern lobe is not reliably detected by the
1415 MHz map.

The 330 to 74 MHz spectral index at the core is $\alpha \sim -0.48 \pm
0.03$.  In the northern direction the spectral index decreases
gradually to $\alpha \sim -1.00 \pm 0.03$ along the arm out to
$125\arcsec$, and reaches $\alpha \sim -1.20 \pm 0.04$ in the northern
lobe.  The apparent small scale structure in the northern lobe is
mostly due to low-level doconvolution artifacts in the 74 MHz image.
On the southern side, the spectral index reaches $\alpha \sim -1.00
\pm 0.03$ at a distance of $65\arcsec$ along the jets, and continues to
decrease into the southern lobe, finally reaching $\alpha \la -1.5
\pm 0.1$.

\section{Discussion}

\begin{figure}
\plotone{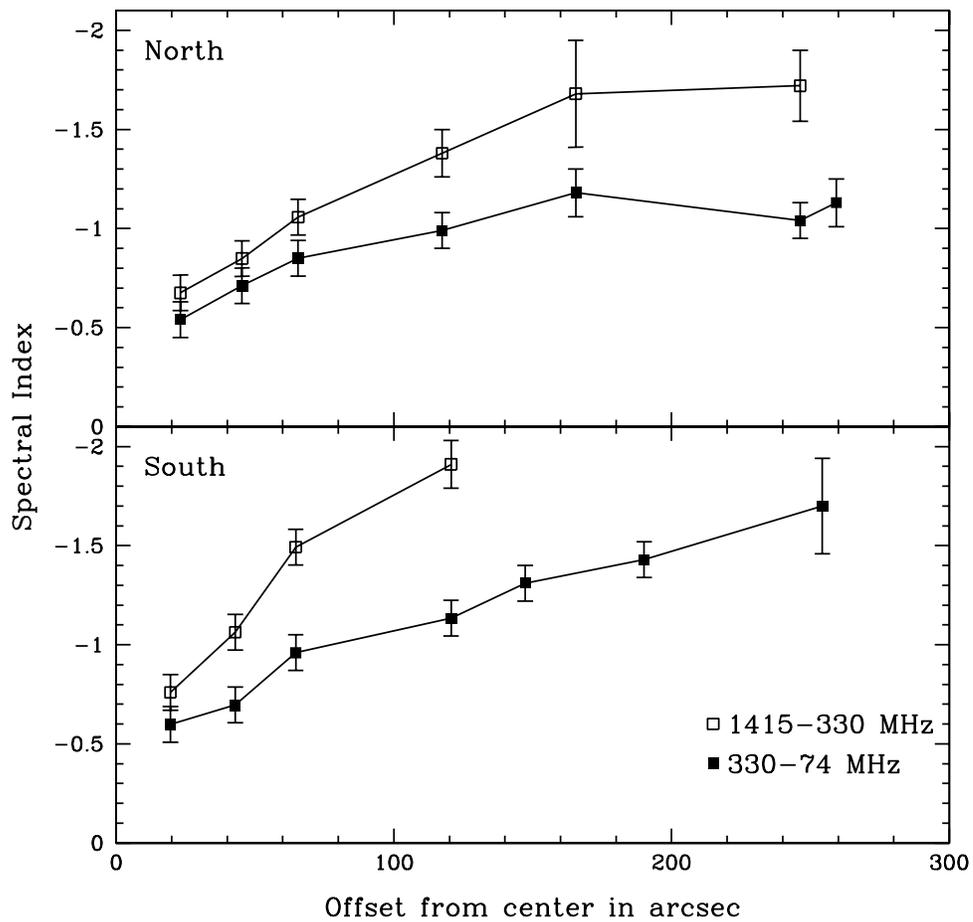}
\caption[lane.fig5.eps]{\small Comparison of $\alpha^{1415}_{330}$ and
$\alpha^{330}_{74}$ along the northern (top) and southern (bottom)
jets of Hydra A.  The distances on the x-axis are measured along the
jets and relative to the center of the source.  Errorbars are
$3\sigma$. \label{fig:SIcomp}}
\end{figure}

There are clear differences between the lower frequency (330 to 74
MHz) spectral index and that at the higher frequency (330 to 1415 MHz)
at a given position within Hydra A.  In Figure ~\ref{fig:SIcomp} we
plot a few representative positions along the northern and southern
jets.  The difference in spectral index is small but consistent, and
significant at greater than a $3\sigma$ level.  The spectral curvature
follows an expected pattern (e.g., Parma et al. 2002); the spectral
slope is steeper at higher frequencies due to aging of the
relativistic particle spectrum over time.

On the other hand the spectral indices calculated from the total flux
at each of our three frequencies indicate that the entire source has a
spectral slope only slightly steeper than the ``typical''
extragalactic source spectral index of $\alpha = -0.7$ (Bridle \&
Perley 1984); we find $\alpha^{330}_{74} = -0.83 \pm 0.03$ and
$\alpha^{1415}_{330} = -0.89 \pm 0.03$ (errors are $1\sigma$ and we
assume a $3\%$ flux calibration precision at each frequency).  Within
the errors, there is little change in the spectral slope over this
frequency range.  These numbers are also consistent with previous
estimates for spectral index in this source (eg. $\alpha^{5 GHz}_{2.7
GHz} = -0.88 \pm 0.08$; K\"uhr et al. 1981).  Using
$\alpha^{300}_{74}=-0.83$ to interpolate the 74 MHz total flux to 178
MHz, we find that Hydra A has $P_{178} = 1.6 \times 10^{26}$ W
Hz$^{-1}$ sr$^{-1}$.  This is in good agreement with previous
estimates for this source (Taylor 1996).

Finally we note that the northern lobe has a flatter spectral slope
than the southern lobe at a given separation from the nucleus.  This
is consistent with it being both the brighter side of the source and
the less depolarized of the two lobes (Taylor 1996; Garrington,
Conway, \& Leahy, 1991; Liu \& Pooley, 1991).  It seems probable that
the southern emission is turning away from us while the northern
emission bends towards us.  Thus the geometry and spectral indices
support the conclusion that Hydra A is inclined to the plane of the
sky, with the northern side lieing closer to us (Taylor 1996).

The northern jet exhibits a turn of about 90 degrees before it expands
into a diffuse lobe, and the jet and lobe are clearly separated in the
image.  In contrast, the radio emission to the south appears to be
bent around by nearly 180 degrees before it finally escapes to the
East in a diffuse lobe.  The steeper spectral slope in the southern
jet could be caused by a superposition of the jet and lobe emission
along the line of sight.  It may also be related to the greater
confinement of the radio source to the south.

\section{Conclusions}

We present moderate resolution VLA maps of Hydra A at 74, 330 and 1415
MHz.  These new images reveal a complicated extended structure, only
hinted at in previously published maps for this powerful FR I radio
source.  Nulsen et al. (2002) show that the cool, X-ray emitting gas
in the central region of Hydra A extends beyond the 6 cm radio
contours. A comparison to our new low frequency observations of Hydra
A shows that these cool regions roughly align with the larger scale
radio emission.  Given the shape and well-defined border of the
northern lobe, we suggest that it may in fact be filling a bubble in
the X-ray gas, analogous to buoyant structures inferred to exist in
clusters such as Perseus A (Fabian et al. 2002).  If so, it would be
the largest example of such a feature identified to date.  A detailed
comparison with the X-ray observations would be required to
investigate this idea.

\acknowledgements

Basic research in astronomy at the Naval Research Laboratory is funded
by the Office of Naval Research.  T. E. C. was supported in part by
the National Aeronautics and Space Administration through $Chandra$
Award Number GO2-3160X, issued by the $Chandra$ X-ray Observatory
Center, which is operated by the Smithsonian Astrophysical Observatory
for and on behalf of NASA under contract NAS8-39073.

\null\clearpage

\onecolumn

\newpage

\begin{table}
\begin{center}
\caption[ObsTable]{Summary of Radio Observations \label{tab:Obs}}
\begin{tabular}[c]{*{5}{l}}
\tableline \tableline
Date & Configuration &  Frequency & Bandwidth &  Integration \\
~~ & ~~ & (MHz) & (MHz) & (min) \\
\tableline 
1988 Aug 14 & D & 1464.9 & 50 & 28 \\
~~ & D & 1514.9 & 50 & 28 \\
1989 Aug 29 & C & 1415.0 & 12.5 & 80  \\
~~ & C & 1430.0 & 12.5 & 80 \\
~~ & C & 1515.0 & 12.5 & 79 \\
~~ & C & 1530.0 & 12.5 & 79 \\
1989 Oct 1 & C & 1415.0 & 12.5 & 79 \\
~~ & C & 1430.0 & 12.5 & 79 \\
~~ & C & 1515.0 & 12.5 & 79 \\
~~ & C & 1530.0 & 12.5 & 79 \\
1990 Mar 31 & A & 320.0 & 6.25 & 36 \\
1990 Mar 31 & A & 330.0 & 6.25 & 36 \\
1990 Jul 28 & B & 330.0 & 6.25 & 39  \\
1992 May 1 & C & 332.9 & 3.125 & 24 \\
1992 Oct 24 & A & 332.9 & 3.125 & 147 \\
1993 Feb 27 & B & 332.9 & 3.125 & 36 \\
1993 Jul 8 & C & 332.9 & 3.125 & 57 \\
1995 Jul 1 & A & 333.0 & 3.125 & 20 \\
1995 Aug 6 & A & 332.9 & 3.125 & 45 \\
1995 Oct 15 & B & 332.9 & 3.125 & 64 \\
1998 Mar 8 & A & 327.5 & 3.125  & 7 \\
1998 Oct 4 & B & 327.5 & 3.125  & 15 \\
1998 Dec 4 & C & 327.5 & 3.125 & 10 \\
2002 Apr 29 & A & 73.8 & 1.56 & 252 \\
\tableline
\end{tabular}
\end{center}
\end{table}

\end{document}